# The Effect of Mechanical Cold Work on the Magnetic Flux Expulsion of Niobium

S. Posen[1], S. K. Chandrasekaran, A. C. Crawford, A. Grassellino, O. Melnychuk, A. Romanenko, D. Sergatskov, Y. Trenikhina

*Fermi National Accelerator Laboratory, Batavia, Illinois, 60510, USA.*

**Abstract**
Expulsion of ambient flux has been shown to be crucial to obtain high quality factors in bulk niobium SRF cavities. However, there remain many questions as to what properties of the niobium material determine its flux expulsion behavior. In this paper, we present first results from a new study of two cavities that were specially fabricated to study flux expulsion. Both cavities were made from large grain ingot niobium slices, one of which had its slices rolled prior to fabrication, and none these slices were annealed prior to measurement. Expulsion measurements indicate that a dense network of grain boundaries is not necessary for a cavity to have near-complete flux trapping behavior up to large thermal gradients. The results also contribute to a body of evidence that cold work is a strong determinant of flux expulsion behavior in SRF-grade niobium.

**Introduction**
When a superconducting radiofrequency (SRF) cavity made of niobium is cooled through its superconducting transition temperature in a background magnetic field, the amount of flux trapped depends on the thermal gradient during cooldown [1] and the history of the material [2]. This has a substantial impact on the quality factor: the impact was large enough in the case of production of cavities for the main linac of LCLS-II at SLAC [3] that the project increased the standard furnace treatment temperatures to enhance expulsion behavior [4]. Different minimum heat treatment temperatures >800 C were required to produce strong expulsion behavior for different production groups of niobium used in the project, but it isn't understood why certain groups require hotter temperatures than others. For future production runs of cavities requiring high quality factors, such as a high energy upgrade to LCLS-II and the PIP-II project at Fermilab, it is important to understand the properties of the material that determine expulsion behavior so that material specifications can be improved.

The expulsion behavior observed may be explained by a competition between two forces. The first would be a force that increases with increasing thermal gradient, which tends to expel flux out of the material. The second would be a pinning force, the overall influence of which would be affected by pinning sites in the material. But what are the dominant pinning sites in SRF-grade niobium? Some proposals for relevant quantities that could influence pinning include impurity concentration, density of grain boundaries (see correlation between improved expulsion and grain growth in [2]), dislocations (see degradation of flux expulsion behavior with cavity deformation in [5]), and hydrides [6].

To perform a focused study on one of these factors, an experimental program was developed, the first results of which are reported here. The goal was to fabricate two cavities with nearly identical material and grain growth density, but very different amounts of cold work. Special precautions were taken during the fabrication of these cavities, including not annealing the material at any point between slicing the material from the ingot and the first expulsion measurements. This fabrication process is described in the next section.

**Cavity Fabrication**
Four slices of niobium were cut via wire EDM from an ingot produced by CBMM in 2005. Two of the slices were cut to 2.8 mm thickness (one is shown in Figure 1) and two were cut to 5 mm thickness. All four slices were treated with

---
[1] email: sposen@fnal.gov

BCP and the 2.8 mm sheets were directly sent forward to cavity production.

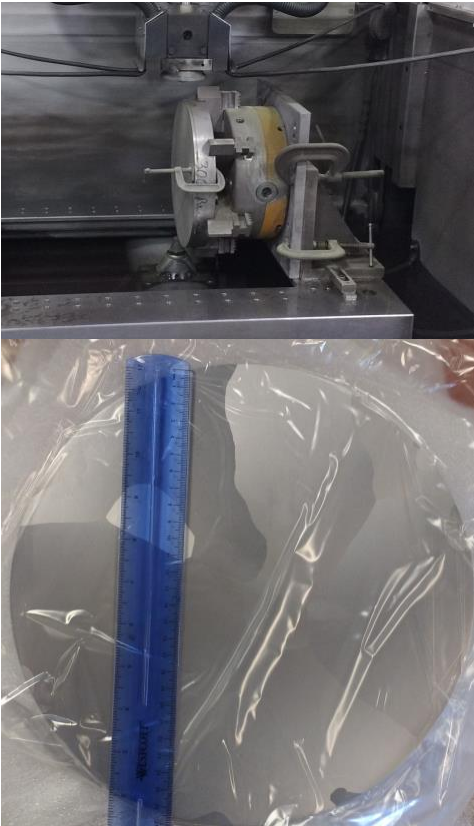

*Figure 1: Wire EDM cutting of niobium ingot (top) and BCP'd slice showing grain structure (bottom).*

The 5 mm sheets were sent to ATI Specialty Alloys and Components, where they were cold rolled to 2.8 mm thickness. Each disc was rolled 50% of the total reduction in one direction (0°) then turned 90° and cross-rolled the remaining 50%. After rolling, the discs had a significant saddle-shaped curvature to them, so they were then subjected to 2-3 rounds of roller levelling. Even after this flattening, process, the final discs were 1-2 cm out of flat (measured as the maximum height of the disc when lying on a flat surface). The diameter of the rolled discs had significantly increased as a result of the thickness reduction. They were waterjet cut to the same diameter as the non-rolled discs, 11 inches. The final product is shown in Figure 2.

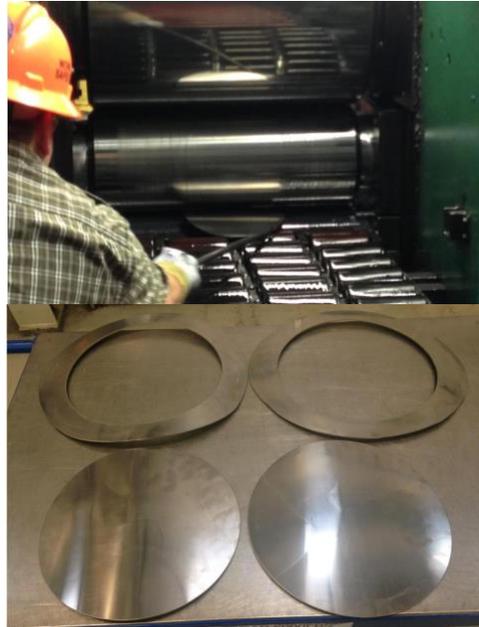

*Figure 2: Rolling process at ATI (top) and final discs with cutoffs (bottom).*

All four sheets were then sent to RI Research Instruments GmbH for cavity fabrication, starting with deep drawing the sheets into the half-cell shape. It should be noted that *the material was not annealed at any point between slicing from the ingot and deep drawing or at any point in the process*. The two non-rolled sheets were formed successfully (even though neither had large central grains as is advisable for large grain material [7]); however, both of the rolled sheets tore near the irises, as shown in Figure 3. This tearing is an indication that the process of modification of the intra-granular crystal lattice by means of cold work was successful.

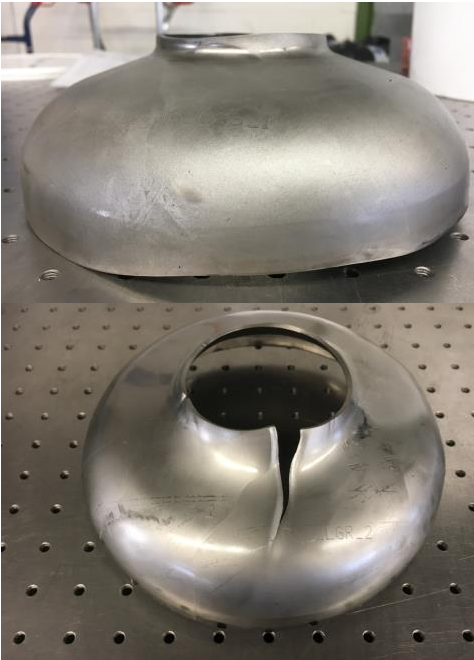

*Figure 3: The non-rolled material was formed into half-cells without tearing (top), but the rolled material tore near the iris (bottom).*

The torn half cells were repaired by milling out a region around each tear and fabricating plugs from the extra rolled material (see the annular pieces in Figure 2). The plugs were successfully electron-beam welded into place, as shown in Figure 4 for one of the half-cells. The extra welds are expected to provide some local heating, but not more severe than a standard equator or iris weld. The impact on flux expulsion behavior is expected to be tolerable for the purposes of this experiment.

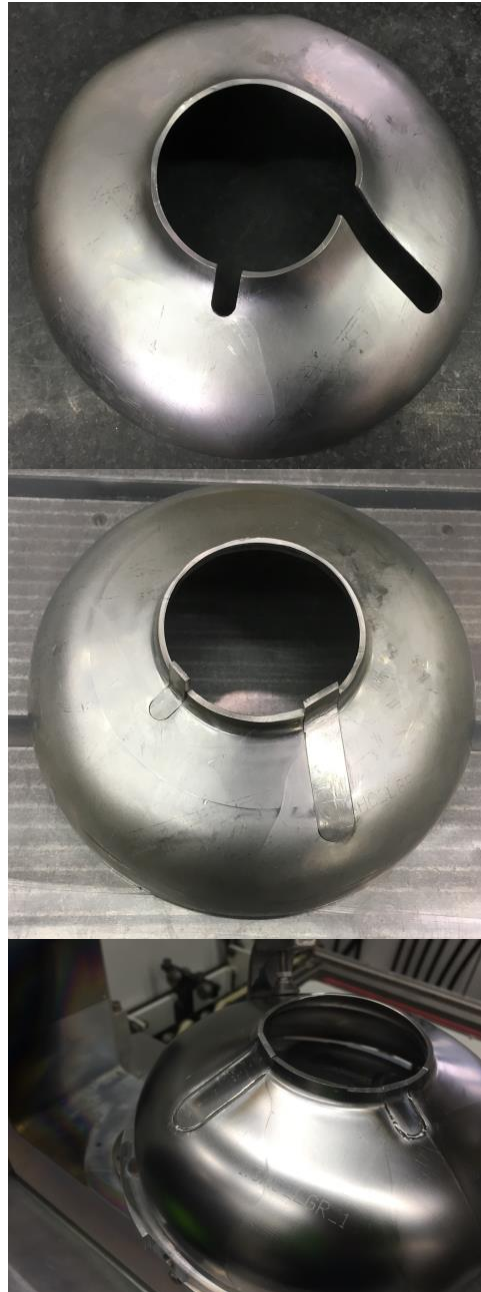

*Figure 4: The repair process for the rolled half-cells involved milling out the torn section (top), inserting plugs of matching material (middle), and welding them into place (bottom).*

The completed cavities are shown in Figure 5.

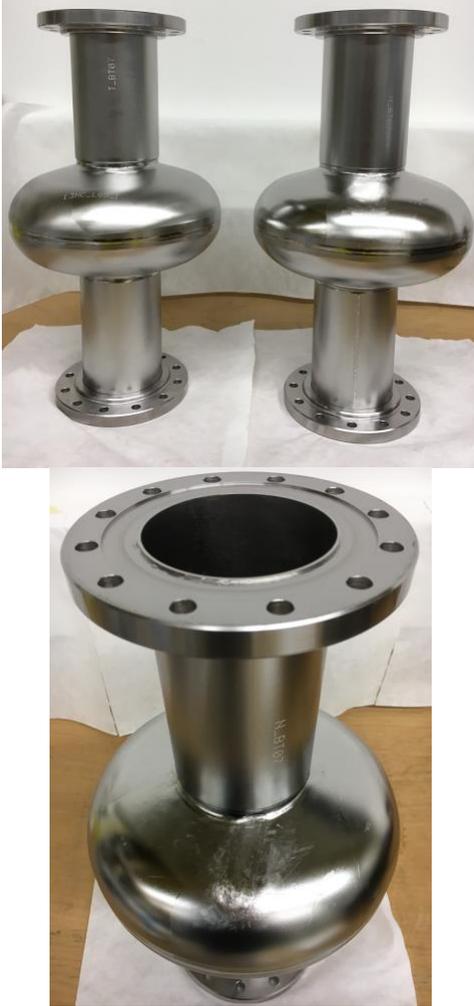

Figure 5: The non-rolled and rolled material cavities together (top) and view of one of the repaired regions in the completed cavity made from rolled material (bottom).

**Measurement of Expulsion**

The cavities were assembled for cryogenic testing immediately after receipt from RI without any chemical or heat treatment. They were assembled without RF antennas, and only magnetic measurements were carried out to probe expulsion, using methods reported previously [1]. For this measurement, external coils apply an axial magnetic field, and three fluxgate magnetometers affixed to the cavity equator measure the field (apparatus shown in Figure 6). As the cavity is cooled through the superconducting transition temperature, the magnetic field is observed before ($B_{NC}$) and after ($B_{SC}$) the cavity becomes superconducting. The ratio $B_{SC}/B_{NC}$ is measured as a function of temperature difference $\Delta T$ across the cavity by performing multiple cooldowns. A larger shift in magnetic field represents more complete expulsion, up to a maximum determined by simulation and by measurement while the cavity is fully superconducting (approximately 74% enhancement for these cavities).

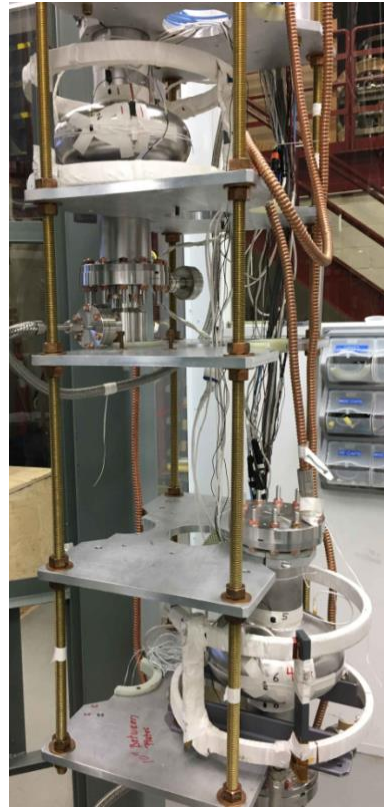

Figure 6: Large grain cavities suspended from the dewar top plate after instrumenting for expulsion measurement.

The cavities were cooled as quickly as possible to minimize the growth of niobium hydrides, which is expected particularly in the range of 100-150 K, as the cavities had not yet undergone the typical hydrogen degassing heat treatment. Each cavity spent less than 12 minutes in this temperature range during cooldown, and neither cavity exceeded 100 K during thermal cycles, as shown in Figure 7. Previous studies suggest that hydrides may play a role in flux trapping [6]. However, the goal of this study was to study effects relevant to state-of-the-art SRF cavities, and the formation of large hydrides is not expected in SRF cavities given standard treatment to avoid Q-disease [8].

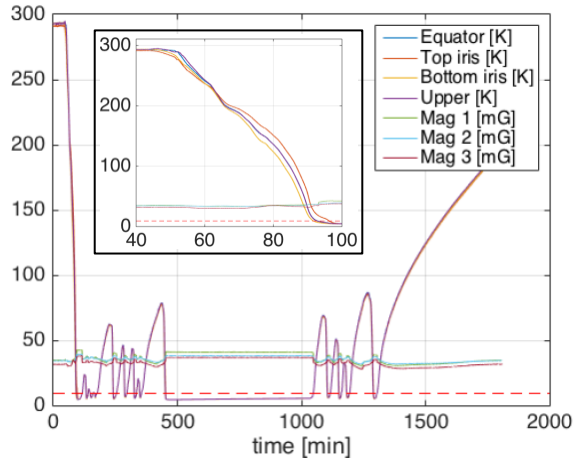

*Figure 7: Thermal cycles performed to measure expulsion in the large grain cavities (rolled cavity shown here). The inset shows detail of the initial fast cooldown from room temperature.*

For each transition through the critical temperature, the expulsion ratio was recorded, and the results are collected in Figure 8.

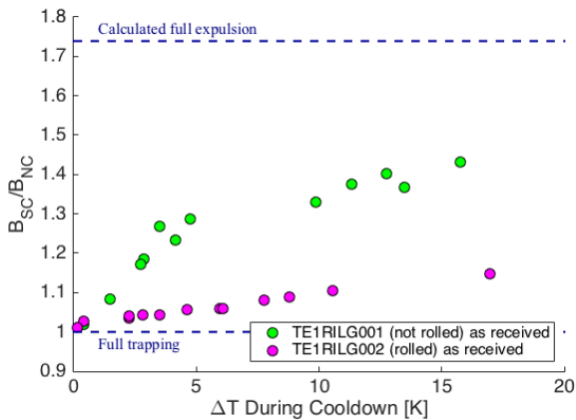

*Figure 8: Expulsion ratio as a function of thermal gradient for the two large grain cavities. The non-rolled cavity has middling expulsion and rolled cavity has nearly full trapping behavior even up to the maximum thermal gradient.*

To supplement the expulsion measurements, microscopy measurements were performed on representative samples. Small samples were cut from the annular section of rolled material and from a non-rolled slice from the same ingot. Electron backscatter diffraction (EBSD) measurements were performed both as a measurement of the dislocation content in the material and as a way to evaluate if subgrain boundaries were formed. The results on the rolled material are presented in Figure 9. The non-rolled material is in progress and will be compared to the rolled material.

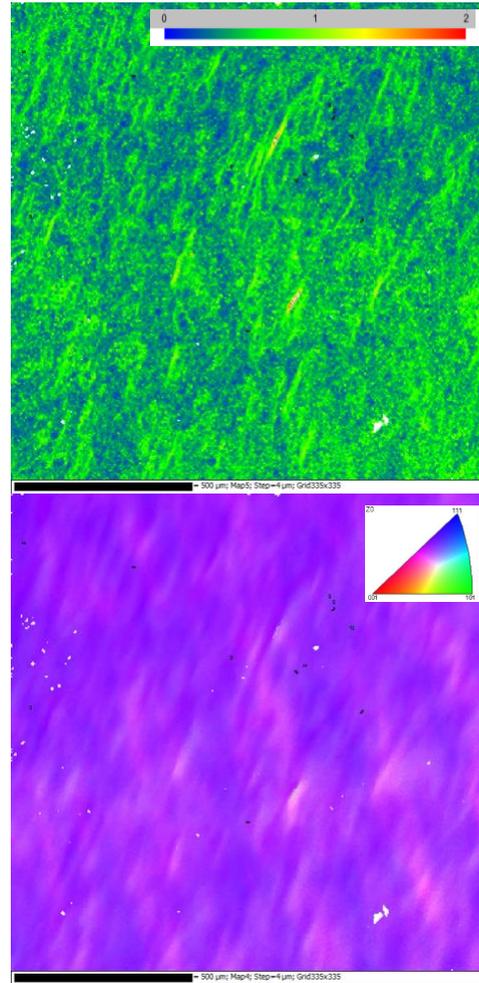

*Figure 9: EBSD Measurement of the rolled material shows significant intragranular misorientation*

**Discussion**

Both cavities show incomplete expulsion even at the largest thermal gradients. The cavity made from the rolled material shows especially weak expulsion behavior, showing near full trapping for the full range of measurement. These results suggest several important implications for the relationship between flux expulsion behavior and the material used to fabricate a cavity:

1. Use of large grain material is not sufficient to ensure flux expulsion—in fact large grain material can have very poor expulsion.
2. The dominant mechanism for flux trapping is not yet fully clear, but it does

not appear to rely on the presence of grain boundaries.
3. Cold work appears to play a substantial role in flux trapping. This is evident from the strong trapping behavior of the cavity made from rolled material.

These results are in good agreement with a previous experiment performed on a fine grain cavity, shown in Figure 10 (results reproduced from [9]). In this experiment, a fine grain cavity with strong expulsion was subjected to deformation—and therefore cold work—by frequency tuning. Light tuning (higher in frequency by 1 MHz then back to the starting frequency) had little effect on expulsion, but heavy tuning (10 MHz) resulted in a significant degradation of expulsion behavior. A 900 C was sufficient to recover the previous expulsion behavior.

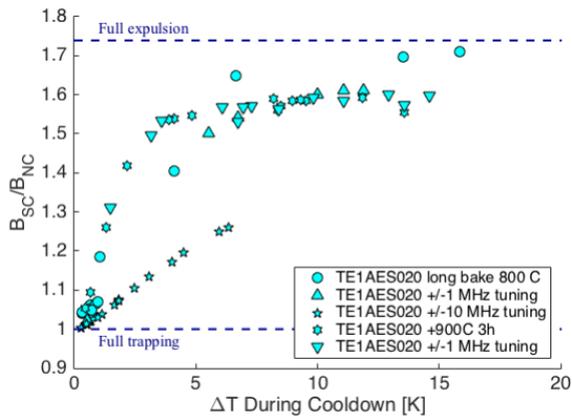

*Figure 10: Substantial deformation (via frequency tuning) of a fine grain cavity degrades expulsion, then a subsequent 900 C bake recovers strong expulsion behavior* [9]

**Conclusion**
Two cavities were made from initially nearly identical large grain niobium material (consecutive slices of the same ingot). One of the cavities had its material rolled to introduce cold work. Tearing occurred during deep drawing of the rolled material, which was subsequently repaired. None of the material was annealed between cutting from the ingot and the first measurement of flux expulsion. The flux expulsion behavior was middling for the non-rolled material and close to full trapping for the rolled material even up to the highest thermal gradients ~18 K. This indicates that cold work, by virtue of dislocations, plays a critical role for flux expulsion behavior, and that grain boundaries may not be important for flux trapping.

These results contribute to a body of evidence (e.g. Figure 10) that cold work in SRF-grade niobium is crucially important to determining the flux trapping behavior. Previous results have demonstrated that hydrides trap flux preferentially compared to grain boundaries [6] and that large grain cavities can have imperfect expulsion even at large thermal gradients [10], but here it is shown for the first time that *near-full trapping* is possible even with a very sparse distribution of grain boundaries over the cavity surface.

The two cavities will next undergo bulk electropolishing to remove the damage layer on the interior surface, then they will be treated in a vacuum furnace at 600 C. Following this, flux expulsion and RF behavior will be measured. The furnace temperature is chosen to degas hydrogen while minimizing recrystallization of the niobium material. The cavities will undergo alternating cycles of cryogenic measurement and heat treatment at increasing temperatures to determine the expulsion behavior as a function of treatment temperature. The strongly different levels of cold work in the two cavities should be helpful to gain insight into the impact of initial cold work on the minimum heat treatment temperature required to reach strong expulsion behavior.

In addition to cavity measurements, microscopy will be carried out on samples cut from the annular section of rolled material as well as samples made from non-rolled material cut from the same ingot. This will include EBSD measurements to evaluate the production of grains/subgrains during rolling and the impact of heat treatment.

**Acknowledgements**
Thanks to Fermilab village machine shop for wire EDM cutting ingot slices and Material Development and Test Lab for carrying out chemical etching. Thanks to ATI Specialty

Alloys and Components for rolling and cutting two of the slices (Figure 2 provided by ATI). Thanks to RI Research Instruments GmbH for fabricating cavities and developing repair method for the rolled material (Figures 3-5 provided by RI). Thanks to Fermilab cavity preparation, testing, and cryogenics teams for contributions to data collection.

This work was supported by the United States Department of Energy, Offices of High Energy Physics. Fermilab is operated by Fermi Research Alliance, LLC under Contract No. DE-AC02-07CH11359 with the United States Department of Energy.

**References**
[1]  A. Romanenko, A. Grassellino, O. Melnychuk, and D. A. Sergatskov, *J. Appl. Phys.*, vol. 115, no. 18, p. 184903, 2014.
[2]  S. Posen *et al.*, *J. Appl. Phys.*, vol. 119, no. 21, p. 213903, 2016.
[3]  J. N. Galayda (ed.), 2015.
[4]  D. Gonnella *et al.*, *Nucl. Instruments Methods Phys. Res. Sect. A Accel. Spectrometers, Detect. Assoc. Equip.*, vol. 883, no. August 2017, pp. 143–150, 2018.
[5]  S. Posen, *Proc. Seventeenth Int. Conf. RF Supercond. Whistler, Canada*, no. tuxba02, 2017.
[6]  J. Köszegi, O. Kugeler, D. Abou-Ras, J. Knobloch, and R. Schäfer, *J. Appl. Phys.*, vol. 122, no. 17, 2017.
[7]  W. Singer *et al.*, *Phys. Rev. Spec. Top. - Accel. Beams*, vol. 16, p. 12003, 2013.
[8]  H. Padamsee, J. Knobloch, and T. Hays, New York: Wiley-VCH, 2008.
[9]  S. Posen, *Proc. Eighteenth Conf. RF Supercond. Lanzhou, China*, 2017.
[10] P. Dhakal, *Proc. TeSLA Technol. Top. Work. Fermilab*, 2017.